# An Integrated Usability Framework for Evaluating Open Government Data Portals: Comparative Analysis of EU and GCC Countries

Fillip Molodtsov, University of Tartu, Estonia

Anastasija Nikiforova[1], University of Tartu, Institute of Computer Science, Estonia

**Abstract**: This study explores the critical role of open government data (OGD) portals in fostering transparency and collaboration between diverse stakeholders. Recognizing the challenges of usability, communication with diverse populations, and strategic value creation, this paper develops an integrated framework for evaluating OGD portal effectiveness that accommodates user diversity (regardless of their data literacy and language), evaluates collaboration and participation, and the ability of users to explore and understand the data provided through them. The framework is validated by applying it to 33 national portals across European Union and Gulf Cooperation Council (GCC) countries, as a result of which we rank OGD portals, identify some good practices that lower-performing portals can learn from, and common shortcomings. Notably, the study unveils the competitive and innovative nature of GCC OGD portals, pinpointing specific improvement areas such as multilingual support and data understandability. The findings underscore the growing trend of exposing data quality metrics and advocate for enhanced two-way communication channels between users and portal representatives. Overall, the study contributes to accelerating the development of user-friendly, collaborative, and sustainable OGD portals while addressing gaps identified in previous research.
**Keywords:** *Open data portal, Open data, Open government data portal, Open government data, OGD portal, Framework, Usability, Sustainability, Open data ecosystem, European Union, GCC, Gulf Cooperation Council*

## 1 INTRODUCTION

A fundamental principle of open government is transparency of government information and operations, which serves as the basis for its two pillars: *participation* and *collaboration* [39]. Open data and Open government data (OGD) are a demonstration of transparency, which can create new opportunities for participation and interaction with government and offers new grounds for collaboration between diverse stakeholders [52]. Open government data is the practice of making government data available to the public for use in response to public demand for access and reuse

---

[1] nikiforova.anastasija@gmail.com

[60]. Municipal, state, federal, and national entities are becoming data publishers. Users, incl. citizens, businesses, esp. Small and Medium-sized Enterprises (SME), media (journalists) participating in OGD (movement) create applications, services, maps and other OGD re-use outputs. It also generates improved data, articles and news, turning government data into information and actionable insights, often of social and economic value [12, 41, 51].

The European Commission's Open Data Maturity Report (ODM report) [40] found that 13 countries (almost half of EU members) scored 90% or above on the portal dimension, with France receiving the highest score. However, a high ranking does not imply perfection, but only relative quality. ODM reports, however, are based on self-reports provided by representatives of the OGD initiative, which suggests that the user's perspective may be omitted from its assessment as is not the purpose of this index. To accelerate efforts to develop user-friendly, collaborative, robust, and sustainable portals, it is essential to identify the current state of the art and the best practices that those portals should adhere to, taking into account current trends both in the field of open data and more broadly. This dynamic of continuous development leads to the fact that many developed frameworks become irrelevant or limited, and the evaluation of portals using them and the further development of an agenda to improve its quality prevents the implementation of a sustainable portal that would meet the needs and expectations of users. Many existing indexes and benchmarks have been proposed in the literature to evaluate OGD efforts, however, [19] that studied many of them, suggests that future research should try to integrate the different frameworks, since although the specific target area of OGD benchmark could be performed separately, it would be useful to have an integrated framework. In addition, it is emphasized that benchmarks with a large number of geographic regions are needed. In this study, we address both gaps by answering the raised call.

This study develops an integrated framework for evaluating the usability of an open data portal. This framework is based on previous frameworks identified through the Systematic Literature Review (SLR), supplemented by annotations obtained from portal examinations (desk research). The developed framework is tested on a sample of national portals of the European Union (EU) and Gulf Cooperation Council (GCC) countries. Applying the framework to a sample of 33 countries allows not to test the framework, and to conduct a preliminary analysis of portals constituting the sample. The framework ranks OGD portals, identifies the most competitive portals, good practices that lower-performing portals can learn from, and common shortcomings. Ultimately, the goal of the OGD initiative is to assist governments in becoming more responsible and transparent in storing, accessing, analyzing, and sharing data. Portals that are more user-friendly and easier to use, potentially increase user engagement. Greater engagement means closer interaction between data sources, producers, and consumers. Greater and closer interaction and collaboration results into richer portal content, which starts the cycle of open government (data) success, thereby contributing to the maturity, resilience, and sustainability of the public / open data ecosystem.

The EU and GCC national OGD portals were chosen as the sample as, according to [30], cooperation between the EU and the GCC countries is crucial to better achieve their political and economic goals.

It is important to compare OGD portals from these two regions and determine whether they are comparable and can cooperate ensuring cross-border interoperability and thereby (indirectly) strengthen political and economic ties. In addition, GCC OGD portals are understudied.

The rest of the paper is structured as follows: Section 2 provides the research methodology, Section 3 presents the results of the SLR, Section 4 presents the developed framework, Section 5 presents results of applying the framework to selected portals, and Section 6 concludes the paper.

## 2 METHODOLOGY

The integrated portal usability framework for evaluating OGD portals is developed based on a systematic literature review (SLR), whose findings are combined with the experience gained from examining top OGD portals, and comments from leading experts on the principles of portal design [16, 29, 41–43]. The developed framework is then tested on a sample of 33 national OGD portals with 27 EU portals and 6 GCC portals. As a result: (1) each individual portal is assessed, the scores are summed up for each dimension and a total score is calculated; (2) portal ranking is conducted; (3) the top portals (best performers) are determined for each criterion / dimension, deriving the best practices observed on national portals, which is part of the qualitative analysis carried out alongside the quantitative analysis; (4) trends in the design of portals are determined and collaborative initiatives between portals are identified.

First, the SLR is carried out, for which a 5-step process for analyzing previous research is developed based on systematic review procedures [20], consisting of (1) question definition, (2) study selection, (3) study relevance and quality assessment, (4) data extraction, (5) data synthesis.

### 2.1 Question definition

To achieve the set objective, the following questions were defined established:
- *Q1*: Open data portals of which countries or regions were analyzed by the previous research?
- *Q2*: What OGD portal assessment frameworks, guidelines, or feature lists have been used in previous studies?
- *Q3*: What usability features were identified as lacking when evaluating OGD portals by previous research?
- *Q4*: What recommendations or strategies have been proposed to improve the usability and accessibility of OGD portals in previous studies?

Q1 is intended to assess which countries or regions have the most researched open data portals. It also aims to establish whether any countries or regions are doing significantly better or worse compared to others. Q2 is intended to provide an overview of the criteria that should be considered when evaluating OGD portals in addition to the portal features discussed previously. Q2 aims to identify a list of candidate frameworks to be used in developing an integrated framework. Q2, Q3 provides insights of the common weak points of the evaluated portals, while Q3 and Q4 aims to derive proposals for improving OGD portals, both of which subsequently become part of the developed integrated system under development.

## 2.2 Study selection

To identify relevant literature, digital libraries covered by Web of Science and Scopus were used. These digital libraries were selected because they provide comprehensive coverage of high-quality peer-reviewed scholarly articles and research papers, which is essential for rigorous academic research. The search includes articles, conference papers, and book chapters published in English in the last six years, as the topic is dynamic, where new works will cite the most influential prior works. These databases were queried for the keywords *(("open data" OR "open government data") AND portal ) AND (usability OR evaluation OR assessment OR "user-cent*" OR analy\* OR quality))*, while the search cope was title, abstract, and keywords. As a result, 108 and 176 articles were found in Scopus and Web of Science, respectively.

## 2.3 Study relevance and quality assessment

Scopus and Web of Science results were merged, and duplicates were identified and eliminated. Papers without access to their full texts were removed. The title and abstract were then scanned to determine the relevance of the study. The following criteria and corresponding grading were defined: (1) the study that focuses on assessing the usability of open data portals receives a grade of 1 (the most relevant); (2) the study that mentions the usability assessment of open data portals receives a grade of 2; (3) the study that mentions open data portals receives a grade of 3; (4) the study that is not associated with open data portals receives a grade of 4. Consequently, articles that received grade of 3 and 4 were filtered out. 82 studies remained for further analysis (Figure 1).

## 2.4 Data extraction

To attain the objective of this SLR, a protocol was developed that collects data on each selected study, including (1) descriptive information, (2) information related to study approach and research design, (3) information related to its quality and relevance, and (4) OGD portal assessment-related information. The developed SLR protocol was adapted from [38, 59] with necessary modifications made to align it with the unique requirements of this research. Descriptive information includes generic metadata and the relevance of the article to this investigation. The approach and research design part gathers the research objective. The quality and relevance part notes any additional quality concerns after reading the paper and checking whether the assessment of the OGD portal is the primary purpose of the study. The OGD portal assessment part of protocol collects answers to the literature review questions defined in section 2.1. Results are made public to support open science and ensure replicability of the study through *[LINK to be added upon acceptance]*.

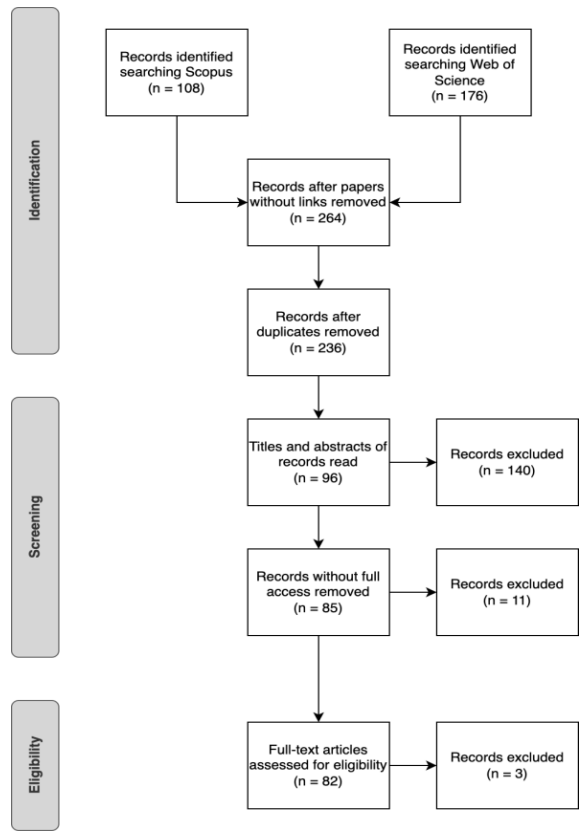

Figure 1: Study selection, relevance and quality assessment (presented using the PRISMA flow diagram).

## 2.5 Data synthesis

The final step of the SLR is data synthesis. We systematically analysed the raw data obtained through the above procedure, the main results of which are presented in section 3. As part of this step, portal features related to user interaction and experience were extracted from 21 articles identified through the data extraction process as most relevant (grade 1 and 2) to build the framework upon (see Table 1). In addition, a list of commonly overlooked usability features has been collected and categorized, along with more non-trivial suggestions for improving OGD portal usability that go beyond refining missed features.

**Table 1:** Studies used for creating the framework.

| Relevance | References |
|---|---|
| High (grade 1) | [25], [32], [22], [40], [48], [4], [1] |
| Medium (grade 2) | [35], [49], [33], [10], [54], [13], [15], [11], [45], [6], [54], [57], [50], [3] |

The procedure used to design it consisted of several steps. First, one of the authors of the manuscript individually reviewed the information collected about each of the selected articles. By analysing the data derived for each metadata dimension for the selected articles, the two authors derived "patterns" that stands for sub-dimensions and dimensions. These were discussed and prioritised between the two authors, and then one author took the lead in developing the draft framework. Both two authors then reviewed the framework, its elements/sub-dimensions, and the dimensions to which they were assigned, suggesting changes. All suggestions were then incorporated into a revised version of the framework, and the framework was reviewed again, resulting in the creation of the final version of the framework.

## 3 RESULTS OF THE LITERATURE REVIEW

2021 witnessed the highest volume of articles published, most of which discussed significant pushpower to make data accessible to address common global challenges such as COVID-19 [14, 23, 35]. This experience reinforced our conviction that it is a good idea to (1) create a composite/ integrated framework compliant with the recent trends, (2) re-evaluate the portals to trigger changes according to their pain points, (3) conduct a cross-border and cross-regional evaluation [8, 9].

### 3.1 Regions and countries covered (Q1)

Forty-seven (47) studies examined portals in the European Union (e.g., [4, 26]). Asia ranks second with twenty-five (25) studies (e.g., [31, 54]), followed by North America with twenty-four (24) studies (e.g., [31, 58]). South America received the fewest mentions (13) (e.g., [6, 33]), followed by Africa (14) (e.g., [3, 33]) and the Pacific (14) (e.g., [11, 32]). It is important to mention that these regions' government portals are part of a larger sample, i.e., it is uncommon for a study to cover portals from a single region. Research suggests that the most potent OGD portals represent countries of Europe, North America, East Asia, and Australasia [31, 32, 37].

### 3.2 Existing frameworks (Q2)

The literature presents a wide range of frameworks, some of which are conceptual and high-level [24], while others are very detailed and straightforward to use [25, 33]. The Open Data Maturity (ODM) reports (also ODMR), published annually by the European Commission (e.g., [22, 40]), are a series of reports demonstrating how frameworks adapt to new trends, adding and removing criteria as they progress. Compared to the 2020 report, the 2022 report adds documentation, API availability, and HVD promotions. Regrettably, certain important elements were omitted from the 2022 report, including defining a sustainability strategy, providing source code, conducting performance index dashboards, and satisfaction surveys. The frameworks are conceptually similar, but they are difficult to compare, which is confirmed by ([5, 19, 27, 28, 52]). For this reason, scholars have begun to reuse frameworks that have been shown to be robust, fit their own purposes, or are easy to apply to their own cases [31]. The usability framework [33] is found the most frequently reused framework. Being among one of the oldest on the list (7 years old), it neglects significant new trends. The transparency-by-design framework [25], which is a revisited/ updated version of [33], covers more aspects and the latest trends, making it more suitable for adopting. However, the framework remains flawed by focusing on transparency rather than usability, especially considering current trends in the resilience and sustainability aspects of OGD portals. For instance, it could benefit from incorporating checks for gamification elements [48], sustainability-related aspects [27], personalization, or more granular multilingualism features. Out of a total of 21 studies analysed, 4 studies presented a framework that was later used in other studies, namely (1) the usability framework [33] reused in 7 studies, (2) "transparency-by-design"-driven OGD portal assessment framework [25] and (3) African OGD portals' media practitioners' preference-centered framework [3] both reused by 2 studies, and (3) "information and system quality framework for Greek OGD sources assessment" [4] reused by 1 study.

## 3.3 Weaknesses of portals (Q3)

As part of Q3, we analyzed and synthesized the notorious missing aspects of OGD portals, grouping them into categories/dimensions. The most commonly mentioned missing **general portal features** are *poor portal navigation*, *lack of prioritization when displaying information* (overloaded with information pages) [31], *multilingualism* [37], and *accessibility features* (e.g., support for screen readers for disabled people). In terms of **data quality**, *metadata absence* [34], *inconsistency* [15], and *data versioning* [22] are among the most severe. *Few valuable datasets* or *generally low data value* (e.g., [35, 36]), and *lack of data visualization and analytical tools* (e.g., [10, 37]) are frequently voiced issues. *Comment sections, forums,* and overall **feedback and support mechanisms** are either missing, or too general, or of poor quality [4, 37]. As for **portal sustainability**, there is *lack of a strategy* [27, 46], *lack of performance dashboards* [44].

## 3.4 Existing recommendations (Q4)

To address the lack of metadata and inconsistency [34] recommends *limiting the number of free form fields for metadata and providing predefined options*, thereby stressing the need for changes in system design. [32] suggest keeping dataset descriptions short and concise, and *keeping the number of actions that need a user account to the minimum*. Some studies stress the need to *use the DCAT-AP vocabulary to make metadata discoverable and understandable worldwide* [21, 57], which at the same time will improve standardization and interoperability. Some suggest migrating portal systems to more advanced technological platforms, such as *CKAN, DKAN, Socrata* data management systems [33, 57]. Some studies emphasize the need to *understand users' needs and demands* [27, 55] and *attract a wider audience* [37]. The study [47] suggested *reducing information pollution on a dataset page, including using algorithms to augment/ enhance metadata management*. To encourage users to use longer search queries, the portal can use *a query recommendation system* and *automatically fill in missing dataset descriptions* [17, 18, 36]. [7, 56] show that *focusing on lay citizens* and allowing them to conduct *searches in their preferred language are both beneficial (multilingualism)* are critical. Last but not least, introducing *gamification elements* and using *storytelling to vulgarize content* is expected to increase the attractiveness and understandability of data portals [48].

## 4 PROPOSED INTEGRATED OGD PORTAL USABILITY FRAMEWORK

The main aspects/dimensions that the proposed integrated OGD portal usability framework focuses on are: (1) inclusivity, ensuring the portal is accessible to a wide range of users, incl. both local/internal and external users of different nationalities and countries being available in different languages; (2) supporting and facilitating user collaboration and active involvement/ participation; and (3) facilitating exploration and understanding of data. The framework consists of nine dimensions, which are divided into 72 sub-dimensions. These dimensions are *(a) multilingualism* (4 sub-dimensions), *(b) navigation* (3), *(c) general performance* (4), *(d) data understandability* (11), *(e) data quality* (9), *(f) data findability* (15), *(g) public engagement* (13), *(h) feedback mechanisms and service quality* (7), and *(i) portal sustainability and collaboration* (6) (see Table 2).

To check the presence of the aspect under consideration, a Boolean assessment (1/0) is predominantly used. However, in the case of accessibility (c4), the web tool / accessibility checker [2] is used, where the portal receives a score of 1 if the tool scores it as 71% or higher (compliance without critical issues). 16 sub-dimensions/aspects (d2-5,7-8,10-11, e1-4,7, and f10-11) are evaluated on a sample basis with a threshold of 70 percent (10 out of 14 datasets) to achieve 1 point. Sample determination works as follows: if the portal supports option to sort datasets by relevance (popularity) and modification date, first four (4) and last three (3) datasets from data catalog list constitute a sample. If only sorting by modification date is implemented, first eight (8) and last six (6) datasets form a sample. If sorting is not implemented - first eight (8) and last six (6) last datasets constitute a sample. The evaluation of the "dataset update frequency accuracy" (e4) checks whether at least 70% of the dataset update frequencies are correct. For example, if the update frequency parameter value is "monthly", the latest modification date should be the current or the previous month. Any modification date will fit as long as the frequency value is "unknown" or "irregular". If the update frequency is specified, but there is no way to verify it (including automatic check indicators), it is not considered fulfilled.

**Table 2:** The proposed framework

| Dimension | Sub-dimension |
|---|---|
| (a) Multilingualism | (1) English is one of the supported languages (*l*), (2) portal interface is available in the supported languages (*m*), (3) portal content is available in the supported languages (*h*), (4) dataset search can be done in English (*h*) |
| (b) Navigation | (1) convenient menu bar structure (*h*), (2) breadcrumb usage (*m*), (3) tabs for content-rich (*m*) |
| (c) General Performance | (1) portal loads in less than 4 seconds (*h*), (2) responsive web design (*l*), (3) no blocking errors or exceptions (*h*), (4) sufficient accessibility level (*m*) |

| Dimension | Sub-dimension |
|---|---|
| (d) Data understandability | (1) HVD promotion (*h*), (2) dataset views (*m*), (3) dataset downloads (*m*), (4) dataset re-use/showcase count (*l*), (5) re-use/showcase display in dataset page (*m*), (6) re-use page dataset list (*h*), (7) data preview (*h*), (8) data visualization (*h*), analytics and filtering tools (*m*), (9) interactive data visualization (*m*), (10) data visualization download (*m*), (11) vulgarized content (described through examples and visual aid) (*h*) |
| (e) Data quality | (1) machine-readable data formats (*h*), (2) basic metadata elements (*h*), (3) update frequency of datasets (*h*), (4) dataset update frequency accuracy (actual vs promised) (*h*), (5) data temporal coverage (*m*), (6) data spatial coverage (*m*), (7) dataset quality rating (*h*), (8) rating explanation (*h*), (9) automated dataset quality checklist (*h*) |
| (f) Data findability | (1) discoverability by publisher (*h*), (2) discoverability by categories (*h*), (3) discoverability by formats (*h*), (4) dataset resource format list catalog preview (*m*), (5) discoverability by tags (*m*), (6) discoverability by license (*l*), (7) sorting by modification date (*h*), (8) sorting by relevance (*h*), (9) sorting by dataset metadata (*m*), (10) dataset tags (*m*), (11) dataset download (*h*), (12) API endpoints (*m*), (13) SPARQL endpoints / RDF files (*m*), (14) recommender system (*h*), (15) featured topics (*h*) |
| (g) Public engagement | (1) possibility to upload use-cases (*h*), (2) present community-sourced / citizen-generated data (*m*), (3) social media support (*h*), (4) notification system (*m*), (5) portal up-to-date information (*h*), (6) shared information about sessions, events created to promote the open data (*h*), (7) personalization features (l), (8) badges (*h*), (9) rewards (*l*), (10) quizzes (*h*), (11) competition (*m*), (12) request forms (*h*), (13) request tracking (*h*) |
| (h) Feedback mechanisms and service quality | (1) portal-wide comment sections or forums (*m*), (2) direct publisher-user communication (*h*), (3) comment sections or forums for datasets (*h*), (4) dataset usefulness assessment (*m*), (5) guidelines (*h*), tutorials, manuals, FAQ (*h*), (6) contact for support (*m*), (7) suggestion for improvement form (*m*) |
| (i) Portal sustainability and collaboration | (1) sustainability strategy available (*m*), (2) performance indexes dashboards or statistics (*h*), (3) regional governments collaboration mentions (*m*), (4) international collaboration mentions (*h*), (5) user satisfaction survey (*h*), (6) code is open source (*l*) |

[a] {*l,m,h*} (in brackets) stands for weights - importance of aspects in relation to the central idea of framework (*low*, *medium*, and *high*).

Despite our best efforts to classify the aspects according to their primary dimension, certain elements may find themselves in different dimensions or even serve a substantial purpose in other dimensions as a side effect. In addition, aspects vary in the level of importance of the central ideas of the framework. Therefore, there is a need for a weighing system. To introduce a weighting system, we first consult the literature to identify potential systems that we could reuse or adopt. One of the most popular approaches is to use equal weights (1) for dimensions and aspects ([25, 33, 57], [52]), (2) for dimensions/aspects ([52, 61]), (3) equal or dimensions but different for aspects within these dimensions ([45, 49], [52]). We do not choose this option because, despite its simplicity, this approach is often criticized even by those who use it (e.g., [26]). Instead, a priority-based option ([52]) is used, where the score of each aspect is multiplied by its importance in relation to the central concepts of the framework. We define three levels of importance: *low, medium*, and *high*, which are mapped to 1, 2, and 3 respectively. The overall portal score is determined by adding the multiplied values, where $\gamma$ is the overall score, $x_l$, $x_m$, $x_h$ - is the score of the aspect marked as low, medium, high importance, respectively.

$$\gamma = \sum(x_l) * 1 + \sum(x_m) * 2 + \sum(x_h) * 3$$

## 5 APPLICATION OF THE DEVELOPED FRAMEWORK: ANALYSIS OF THE RESULTS

The developed integrated OGD portal usability framework was applied to 33 EU and GCC OGD portals, where the individual portal scores were calculated for each dimension, which were then summed to produce a total portal score. Figure 2 presents the portal ranking, in which France is found the leader, which is consistent with the results of previous studies (e.g., [40]). However, the French portal did not receive the maximum score of 176, having some room for improvement. Nevertheless, its score of 138 is objectively high, leaving Saudi Arabia (121) 17 points behind in second place. Eleven national OGD portals passed the 100-point threshold, and only five scored below 50, with OGD portal of Kuwait coming in last place. The low rank of the Kuwaiti OGD portal is due to the fact that Kuwait does not have a national open data portal where OGD are provided through the Kuwaiti government online portal with a dedicated OGD provision section. Let us now discuss the results demonstrated by the selected portals by dimension.

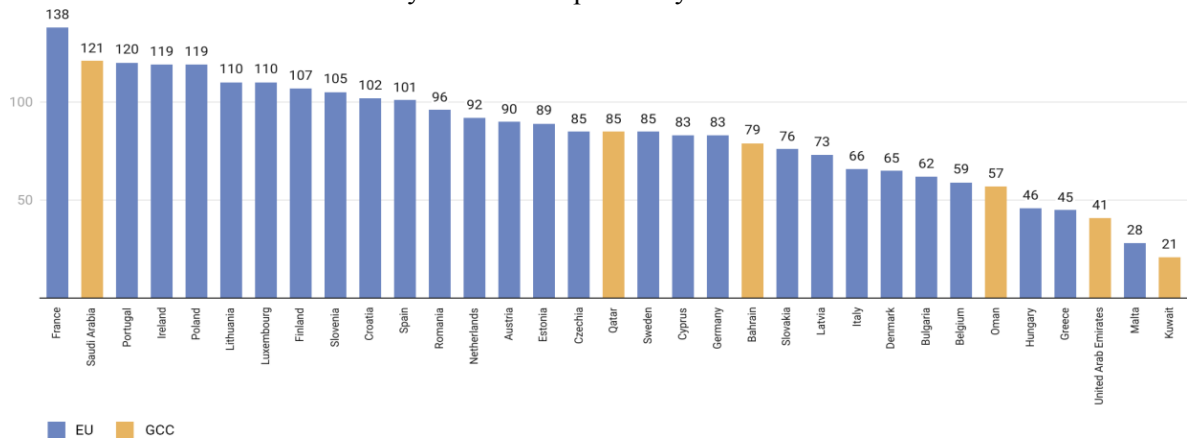

**Figure 2**: Portal ranking.

## 5.1 Multilingualism

Eight portals received the maximum score for the *multilingualism* dimension (9): **Bahrain, Estonia, Ireland, Malta, Oman, Qatar, Saudi Arabia**, and **the United Arab Emirates**. Five of them are GCC states (5 out of 6 GCC countries). The fact that portals GCC provide such a high level of English language support can be interpreted as the willingness of these countries to collaborate beyond the GCC region. The Irish and Maltese portals provide English language support as it is their official language. The Estonian portal, however, demonstrates the successful use of machine-translated metadata.

The Austrian portal only offers to redirect the user to the EU Open Data portal to access machine-translated metadata in the selected language. In addition, publishers can fill out the title and description in English (the second supported language). However, this is not mandatory. The Slovenian portal implements portal-wise translation through the use of Google Translate plug-in, translating content in addition to the user interface (unlike the Hungarian portal). The disadvantage of this approach is that the search still relies on the original metadata of the dataset, which is mainly in Slovenian. German and Italian portals offer exclusively German or Italian interfaces.

Overall, the level of multilingual support is fairly basic among analysed portals and should be improved, with a particular focus on allowing users to search datasets in the language of choice.

## 5.2 Navigation

Fifteen portals (15) received the highest possible score for the navigation dimension, and only the UAE received 0 points. An element that caught our attention (but was not factored into account in the final score) was the inclusion of a site map on the portals of the Czech Republic, Germany, Poland, and Italy. This can serve as an example of good practice that can be considered by other portals to simplify navigation for users (especially first-time users), as well as to provide an understanding of the portal's functionality and overall landscape.

As an example of an improved navigation, all *primary links are placed prominently in the footer* of the Polish portal, facilitating effortless access to each section of the portal. Excellent *usage of tabs and interconnected pages* can be seen on the French portal.

However, the mere existence of navigation elements does not necessarily make them intuitive or practical. *Breadcrumbs* on the Austrian portal may not direct you to previously opened pages. Clicking on the tabs on the dataset page on the Bulgarian portal will take you to a new page without a link back to the dataset's main page. The *priority of some portals' tabs* may seem controversial: on the dataset page, the Portuguese portal displays publisher information prioritized over the dataset metadata although there are publisher-specific pages for that. The UAE portal hides the menu bar on dataset pages, making the content less *consistent* (with the menu bar on pages other than the dataset). Typically, portals provide essential navigation elements, but some of them lack consistency. Additionally, in some cases (e.g., the Irish portal), sections of the menu bar may be flattened or simplified, including adopting good practices.

## 5.3 General Performance

For most of the portals analyzed, the *general performance* is acceptable. The portals of **Poland, Bahrain, Estonia,** and **France** serve as exemplary models of responsiveness and robustness. However, some problems have been identified for some portals. For instance, the Irish portal does not load the last two pages of the oldest / the least recent datasets. The Greek OGD portal requires registration to access data (API tokens to download them), however, when attempting to register, no confirmation was received. In addition, unavailability of dataset resources (files) is common on portals, where only portals of Latvia, Luxembourg, Portugal, Qatar, and Saudi Arabia had resources for all sample datasets. The Maltese portal displays internal identification keys on its pages. The Omani portal has a limit on the number of rows of data that can be downloaded. In general, portals could benefit from more emphasis on assessing the quality of portal functionality and improving page loading speed.

## 5.4 Data understandability

Portals of **Poland, Qatar, France, Saudi Arabia**, and **Bahrain** lead in terms of *data understandability*. Although it is difficult to determine whether a portal is providing *vulgarized content*, within this dimension, *visualization for datasets*, such as those provided for some datasets on the portals of Bahrain, Qatar, Saudi Arabia, Lithuania, Spain, and Poland, can be good examples of making it easier to understand dataset through a visual aid.

The *impact section* and *success stories (also use-cases/showcases/re-uses)* of applications and services built on open data also serve this purpose. Unfortunately, the latter is the most effective and most resource-consuming.

*HVD promotion* takes the form of (a) additional filtering criteria on portals of Ireland, Lithuania, Poland, and Slovenia, (b) promotion on top of the catalog page (Polish portal), (c) featured list (Netherlands portal), or (d) reports highlighting the most valuable datasets (French portal). The Czech portal promotes HVD by holding publishers accountable for the value of their datasets, providing a variety of indicators and dashboards to measure its.

This dimension receives the lowest scores on average. Portals should put more efforts to describe how the data is used (with prior determination of these reuses), what it indicates, and how it can be beneficial. Particular emphasis should be placed on promoting high-value datasets.

## 5.5 Data quality

A common feature of top performers in *data quality* dimension - **France, Czech Republic, Slovenia, Portugal,** and **Croatia**, is the presence of *indicators*, which provides insights into the availability and accessibility of various aspects of the datasets, including resources, specifications, and update frequency accuracy. The Czech portal additionally checks whether dataset contains personal data. The French and Portuguese portals share a common dataset metadata quality indicator that lets users and publishers know if metadata details are missing or incorrect.

Romanian, Slovenian, and Irish portals have *5-star scheme dataset openness rating*, making it easier to determine the openness of datasets. The Bahrain and Qatar portals provides extensive *explanations of dataset schema*. Portal of Qatar allows the *catalog to be downloaded in RDF format* (from the DCAT vocabulary). The Polish portal allows *dataset metadata to be downloaded in CSV and RDF formats*.

The Cyprus portal makes good use of *temporal coverage* parameter. However, the *spatial* parameter may be subject to debate since all data sets use only the value "Cyprus" value is used for all datasets without more granular division. The Finnish portal integrates *spatial coverage into interactive visualizations*.

In general, portals should improve their metadata collection and provision standards to make it richer and more consistent across datasets. The introduction of various *quality indicators* is a trend that should become widespread. Providing a feature that enables the download of dataset metadata should be considered as an additional benefit.

## 5.6 Data findability

The **Irish** portal provides *advanced search capabilities*, while **Portuguese** and **Swedish** portals provide user with *search tips*. The **Finnish** portal provides the ability to *search data within the region selected on the map*.

The **French, Italian, Dutch, Polish**, and **Luxembourgish** portals provide excellent examples of how to implement *"featured topics"* sections. These topics may be general, but for some portals they are quite specific. For example, the Polish portal displays a collection of datasets related to Ukraine, while the French portal provides a rich list of topic-specific featured datasets on topics such as energy, education, culture, COVID-19 etc. Related datasets are displayed on the dataset page in the French and Dutch portals. Unfortunately, not always likeness relation / similarity rate is shown.

In general, portals in this dimension perform adequately. However, we suggest they prioritize exposing API/GraphQL endpoints and making their content accessible, establishing connections between datasets on similarity to facilitate promotion, and highlighting featured topics.

## 5.7 Public engagement

The **Lithuanian** portal is top performer in *public engagement* dimension as it has a lively *news section* full of articles and event announcements. The **Spanish** and **Croatian** portals have rich *report-tracking* features, displaying a list of reports and the status for each of them. In addition, the Spanish portal offers users the opportunity to report a wide range of issues, including data availability, offer suggestions for improvement, and exchange information regarding reuses and initiatives.

Some portals use video content. E.g., the Polish, Czech and Spanish portals have a decent collection of interviews and educational material posted either directly on the portal or on YouTube. The Polish portal has a useful use of *notifications*, allowing users to receive datasets or search results for updates. Examples of *community-sourced datasets* were seen on portals of France and Finland.

On average, the performance of portals in this dimension is quite low. There is not a single example of introducing gamification elements. Personalization in most cases is limited to following and liking datasets. Many portals do not have a list of previously reported issues or the ability of submitting reuse. In addition, social media accounts are rarely active.

### 5.8 Feedback mechanisms and service quality

**Spain, Croatia, Lithuania, France,** and **Portugal** are the top performers in the *feedback mechanisms and service quality* dimension. The *comment section* on the dataset page is a trend implemented by the portals of France, Croatia, and Luxembourg. The usefulness of this feature can be verified by seeing the lively discussion in the corresponding section, where, however, participation of both parties, i.e., not only user, but also a publisher is of importance.

*Dataset usefulness assessment* spanning from voting, 5-score scale to subscribing or following a dataset or page is implemented in some portals, such as Croatian, French, Portuguese, Estonian, Dutch, and Latvian.

Many portals (e.g., French, Portuguese, Austrian, Czech) provide users - potential publishers and users with *guides and manuals*. However, they are often very technical and are unlikely to be understandable to a lay user or tailored to data publishers. Although there are examples (e.g., the Irish, Dutch, and Czech portals) where some manuals are tailored toward lay citizens. A virtual tour of the Polish portal certain help new users easily navigate the platform. The Saudi Arabia portal impresses with its rich variety of *communication channels*: by mail, contact form, address, dataset suggest or request (the difference between both, however, is not clear since the forms are the same), complaint form.

In general, portals should continue to improve their quality of service and feedback functions. To foster communication between publishers and users, portals should create and maintain comment sections. The content of manuals and documentation should be more beginner friendly. There should be forms tailored for different purposes. There are examples (the Austrian, Greek, Bulgarian, and Bahraini portals) where there is no communication with the support service, or it only relies on writing emails. The Belgian case of customer support via online chat is uncommon.

### 5.9 Portal sustainability and collaboration

The maximum score was received by **Finnish, French, Irish, Luxembourgian, Polish,** and **Portuguese** portals. The Polish portal provides the *Open Data Programme for 2021-2027* in the useful material section. The French portal *tracks releases* on a monthly basis, highlighting released datasets and reuses. The German portal celebrated its 10th anniversary in 2023 by sharing past milestones in a blog section.

A comprehensive *user satisfaction survey* is rarely implemented, but Finnish and French portals serve as an example here. Instant satisfaction surveys can be found on many pages of the Saudi Arabian portal. The Germany-Austria-Switzerland-Liechtenstein cooperation is highlighted on portal's strategy pages, demonstrating an example of *international collaboration*, but is not given any attention in the catalogues. The Irish portal promotes regional OGD portal of the Northern Ireland, while on the Qatari portal, the user can build a map based on data sourced from the EU states and Israel.

*The source code of the portals* was found in the associated repositories for 19 of the 33 analysed portals, but few of them were posts links to them on the portal itself. On average, portals perform adequately in this dimension. However, we would suggest conducting more user satisfaction surveys, having a defined strategy, sharing reports, tracking the release of new artifacts (e.g., datasets, visualizations, reuses). The benefits of portal partnerships must be highlighted to promote it within and across regions, as well as to achieve a more cross-border open data ecosystem.

## 6 CONCLUSIONS

This study develops an integrated usability framework for evaluating open data portals that focuses on: (1) the portal's ability to adapt to a diverse user base; (2) promoting user collaboration and participation; and (3) enabling users to understand and explore the data. The framework, consisting of 72 dimensions, was applied to 33 EU and GCC OGD portals to assess their usability. As a result of its application top performers for each criterion / dimension were identified, deriving best practices observed on national portals, and trends in portal design and collaborative initiatives between portals.

The high performance of top European portals (based on EU Open Data Maturity Reports) within this framework can be seen as an indicator of some degree of consistency between the proposed framework and existing widely used benchmarks and indices. However, an alternative view of these portals has been presented based on the unique aspects/dimensions that the presented framework considers. Compared to the EU portals, we found that the portals of Saudi Arabia, Qatar, and Bahrain are competent and even trendsetting in certain sub-dimensions.

Our analysis suggests that portals should focus on improving the multilingual experience, allowing users to communicate their needs to portal representatives or data publishers. It should also make it easier for users to understand how to use datasets and find them on the portal.

It has been observed that there is a growing trend of exposing data quality indicators. Additionally, involving users in the portal ecosystem results in a more vibrant and engaging experience, increasing the likelihood of repeated use of the portal even for newcomers, making them part of the open data ecosystem. A common problem is that portals often fail to adequately highlight implemented features, which are often hidden and difficult to find, not to say mention the lack of features that a user might expect, including assistants, AI-augmented recommender systems, advanced search, or NLP or LLM capabilities for advanced search or examining datasets, as well as gamification elements. In the future, this framework will be also revisited once we have examples of how the above technologies can be advantageous for open data portals. At the moment, it can be rather seen as the "minimum set of requirements" that the OGD portal must comply with.

The results of the analysis indicate that despite the lack of research comparing portals of different regions, incl. the EU and GCC, it is feasible and advisable to conduct a comparative analysis of portals from different regions. While portals from the same region often have similar strengths and weaknesses, studying portals from different regions can provide new insights into how the same features may be implemented differently.

Although cross-border and inter-regional portal collaboration is not widely observed, several examples have been found in both regions. The Qatar portal represents an innovative example of cross-regional innovation, while the German and Austrian portals are promoting their cooperation with the Swiss and Lichtenstein portals, thereby moving towards an enhanced and more sustainable open data ecosystem (among Germany, Austria, Switzerland, and Liechtenstein). While the importance of collaborations in creating a more connected open data ecosystem is often underestimated, these efforts are critical to making progress toward this goal.